\documentclass{ws-procs9x6}

\usepackage{url}

\begin{document}

\title{ASTROFIT:\\
AN INTERFACE PROGRAM FOR EXPLORING COMPLEMENTARITY IN DARK MATTER RESEARCH}

\author{N.NGUYEN$^*$ and D.HORNS}

\address{Institute for Experimental Physics, University of Hamburg,\\
Luruper Chaussee 149, 22761 Hamburg, Germany\\
$^*$E-mail: nelly.nguyen@desy.de}

\author{T.BRINGMANN}

\address{II. Institute for Theoretical Physics, University of Hamburg\\
Luruper Chausee 149, 22761 Hamburg, Germany\\
E-mail: torsten.bringmann@desy.de}

\begin{abstract}
AstroFit is an interface adding astrophysical components to programs for fitting physics beyond the Standard Model (BSM) to experimental data from collider searches. 
The project aims at combining a wide range of experimental results from indirect, direct and collider serarches for Dark Matter (DM) and confronting it with theoretical expectations in various DM models.
Here, we introduce AstroFit and discuss first results.
\end{abstract}


\bodymatter

\section{Introduction}
Various different experiments explore the properties of Dark Matter (DM) while a plethora of theories offer explanations to its nature. It is the goal of the AstroFit project to constrain DM models by combining experimental data from both astrophysics and collider physics, and thus find the best fit regions for the parameter space of the underlying theory. 
AstroFit itself is a Fortran program, serving as an interface between programs used in particle physics for fitting physics beyond the Standard Model, such as Fittino~\cite{fittino1, fittino2, fittino3},
and programs like DarkSUSY~\cite{darksusy, 2darksusy} that can be used to calculate theoretical predictions for direct and indirect detection experiments. An overview of experimental input usable with AstroFit will be given as well as an example of how to use AstroFit in combination with the Fittino program. 
Here, the first results from a Constrained Minimal Supersymmetric Standard Model (CMSSM) fit including information from latest collider (i.e. Large Hadron Collider (LHC)), direct detection and indirect detection instruments are presented. 

\section{Complementarity of Experiments}
\emph{Indirect searches} for Dark Matter concentrate on finding products from DM annihilation, such as photons of different energy ranges (e.g. radio, X-ray, $\gamma$-ray), antiprotons, positrons and neutrinos with specially designed experiments, ranging from ground-based Cherenkov telescopes like H.E.S.S.~\cite{2hess}, MAGIC~\cite{magic} and VERITAS~\cite{veritas} over satellite experiments like Fermi-LAT~\cite{2fermi} and PAMELA~\cite{pamela} to balloon experiments as ATIC~\cite{atic}, among many others. 
For this analysis, we used photon flux upper limits from dwarf galaxies by H.E.S.S.~\cite{hess} and Fermi-LAT as observables, which have been calculated by AstroFit as 

\begin{equation}
\frac{\text{d}\Phi(\Delta \Omega,E_{\gamma})}{\text{d}E_{\gamma}} = \frac{1}{8\pi} \frac{\langle\sigma v \rangle
 }{m^{2}_{\chi}} \frac{\text{d}N_{\gamma}}{\text{d}E_{\gamma}} \times \bar{J}(\Delta\Omega) \Delta\Omega\, , 
 \end{equation}
 where ${\text{d}\Phi(\Delta \Omega,E_{\gamma})}/{\text{d}E_{\gamma}}$ is the differential photon flux, $E_{\gamma}$ the photon energy, $\langle\sigma v \rangle$ the velocity weighted annihilation cross-section and $m_{\chi}$ the mass of the DM particle, ${\text{d}N_{\gamma}}/{\text{d}E_{\gamma}}$ the differential number of gammas produced per annihilation per energy and
$\bar{J}(\Delta\Omega)= ({1}/{\Delta\Omega}) \int_{\Delta\Omega}\text{d}\Omega \int_{l.o.s.} \text{d}l \; \rho^{2}_{\text{DM}} (l)$ the integral over the DM density squared along the line of sight. Instead of using the photon flux upper limit as direct observable, the thermally averaged cross-section times the relative velocity can be used if desired, as some experimental limits are preferably presented in terms of $\langle\sigma v \rangle_\text{max}$.

\emph{Direct detection} instruments such as DAMA/LIBRA\cite{dama}, CoGeNT\cite{cogent} and Xenon \cite{2xenon}, located in underground laboratories, measure signals from DM interactions  with target elements like Xenon, Germanium or compounds such as NaI. Signals are detected via scintillation, phonons or ionization, depending on the experiment. Assuming a scattering of a weakly interacting massive particle (WIMP) as DM candidate with the target material, the spin-independent scattering cross-section per nucleon, conventionally adopted for comparison between experimental results, is calculated as follows in AstroFit: 
\begin{equation}
\sigma^{SI}_{nucleon} = \frac{(Z \sqrt{\sigma_p} \pm (A-Z) \sqrt{\sigma_n})^2}{A^2}\, , 
\end{equation}
with $\sigma_p$ and $\sigma_n$ being the spin-independent cross-section for one proton or neutron, respectively, and Z and A being the atomic and mass number of the target element.
Latest results from direct detection have recently shown a conflict between measurements from different experiments. While the DAMA, CoGeNT and CRESST\cite{2cresst} collaborations each published detections of signals, the Xenon collaboration has shown upper limits lower than the signal regions of the previous experiments for the WIMP-nucleon cross-section. Possible reasons for this diversity are still being discussed. 

As particle physics input from \emph{collider experiments}, information on B- and Z-physics, like masses, edges in mass spectra, widths, asymmetries, etc. have been used in this analysis as well as the anomalous magnetic moment of the muon and latest event rates from the ATLAS experiment at LHC (see \refcite{fittino1, fittino2, fittino3} for details). 
Also, the relic density of Cold Dark Matter (CDM) provided by WMAP~\cite{wmap}, $\Omega_{\text{DM}}h^2 = 0.1123 \pm 0.0035 $
which can be directly compared to the theoretical prediction for thermally produced DM from DarkSUSY, is used as an observable in AstroFit. 

\section{Structure of AstroFit}
As depicted in figure 1, AstroFit provides an extensive database of relevant experimental data to add to a fit process, at the moment implemented in Fittino. While the steering of AstroFit is done by a user-friendly text input file, information on the particle spectrum is given via a SUSY Les Houches Accord (SLHA) file directly from Fittino. From this spectrum file, AstroFit subroutines are designed to calculate the model predictions by using various DarkSUSY functions. The theoretically estimated observables are then compared to actual measurements from astrophysical experiments. From the comparison, a  \(\Delta\chi^2\)-contribution is calculated for each individual observable and is handed to Fittino in each step of the minimization process to be used together with the information from particle physics. In a stand-alone subroutine of AstroFit, the \(\Delta\chi^2\) is calculated by applying $
\Delta \chi^2 = \sum \left(({O_\text{exp} - O_\text{theo}})/{\sigma_\text{exp}} \right) ^2$ for data points or continuously in parabolic increase by extrapolation using the given confidence level for limits or containment regions for claimed signals for any observable. A minimization strategy in Fittino then determines the lowest global \(\chi^2\), and the results comprise the best fit regions for all model parameters. 

 \begin{figure*}[!t] \label{af_fc}
 \centering
\includegraphics[width= .65\textwidth]{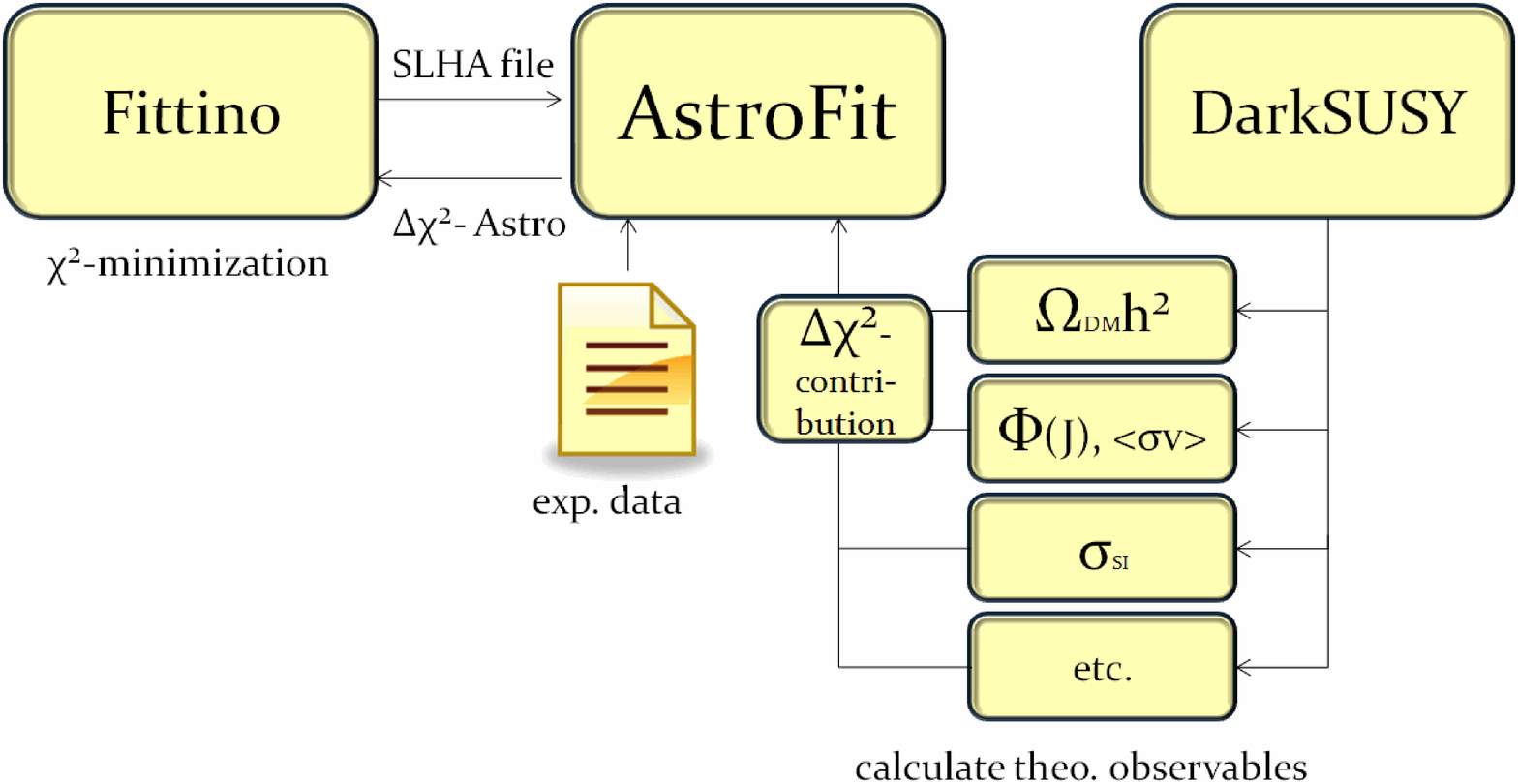} 
   \caption{AstroFit flowchart: Minimization process taking place in Fittino, calculations of theoretical observables using subroutines from DarkSUSY and comparison to astrophysical observables done in AstroFit.}
\end{figure*}

\section{Analysis and Results for CMSSM Model}
Exemplarily, this analysis focussed on the CMSSM, using a Markov Chain Monte Carlo algorithm to  fit the following parameters (defined at the Grand Unification Theory scale): 
$M_0$ --  the universal scalar mass,
$M_{1/2}$ -- the universal gaugino mass,
$A_0$ --  the common trilinear coupling and
$\tan{\beta}$ -- the ratio of the vacuum expectation values of the two Higgs fields.
We assume the neutralino to be the DM particle and set $\mu=+1$.

For our analysis, the data from CoGeNT has been used in one fit and the data from Xenon100 and Xenongoal in two others (compare \refcite{xenon}) . Additionally, fits with data for Xenon1T and CRESST\cite{cresst} are in preparation. 

While claimed DM signals from the CoGeNT experiment could not find agreement with a CMSSM fit, upper limits from the Xenon experiment give constraints in this scenario. 
 In figure 2, fit results are shown using only particle physics information (including latest $2\text{fb}^{-1}$ results from LHC) and the relic density of CDM. Adding further constraints from indirect and direct detection, i.e. photon flux upper limits and results from Xenon100 data, leads to the results shown in figure 3. 
In the fit, the bulk region is already excluded by all other input conditions, while adding Xenon100 limits provides an additional constraint on both the coannihilation and the funnel region.

 Photon flux upper limits do not constrain the CMSSM parameter space so far. However, using the latest joint likelihood results for dwarf spheroidal galaxies from Fermi-LAT (see \refcite{fermi}) could show first effects. 

 \begin{figure*}[!t]
\begin{minipage}[t]{0.48\textwidth}
\includegraphics[width= .9\textwidth]{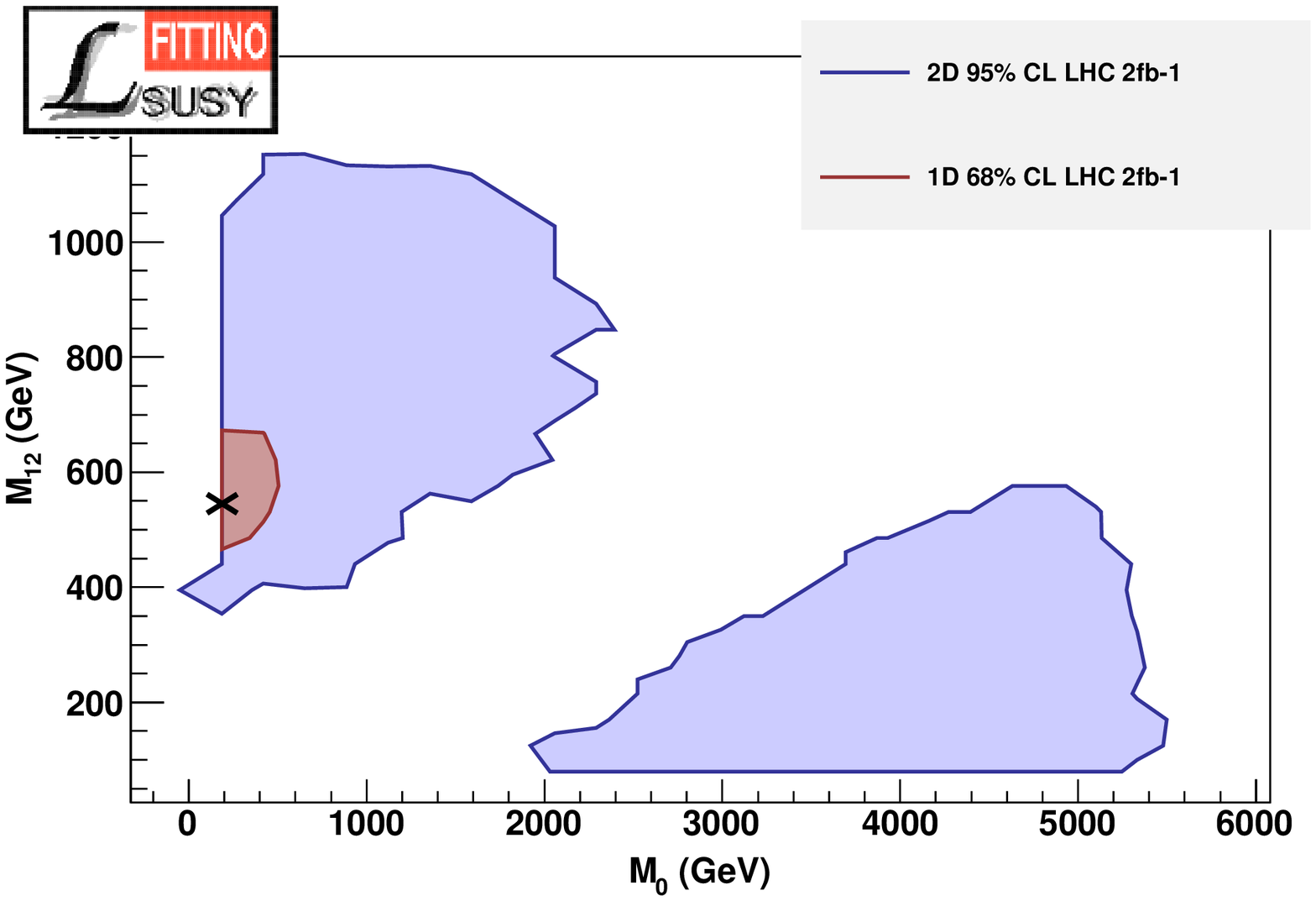} \label{2fbnoAFlogo}
   \caption{$M_{0}$ - $M_{1/2}$ 2D 2$\sigma$ contour region, for a fittino fit using observables from collider production and the relic density of CDM.}
\end{minipage}
\;\;\;
\begin{minipage}[t]{0.48\textwidth}
\includegraphics[width= .9\textwidth]{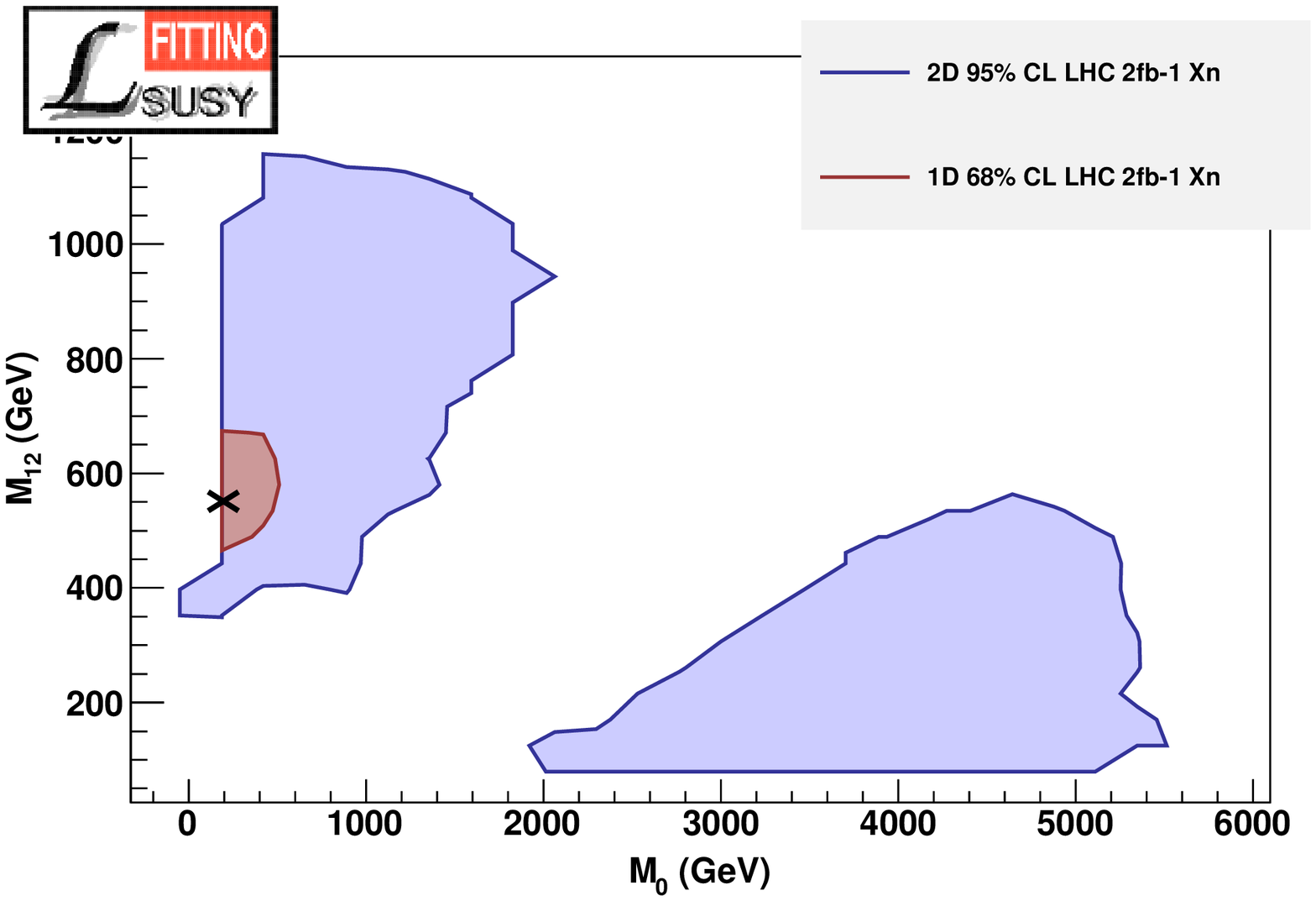}  \label{2fbAFlogo}
   \caption{Same as Fig. 2, including also Xenon100 11d limits. The additionally constrained models lie mostly in the coannihilation and funnel region.}
\end{minipage}
\end{figure*}

\section{Conclusion and Outlook}
Using all available results from DM searches can help confirm, constrain or exclude regions in parameter space of DM models remarkably, making it possible to edge closer to understanding physics beyond the Standard Model in general and the nature of DM in particular. In this study, observables from particle physics have been combined with the relic density of CDM, photon flux upper limits from dwarf spheroidal galaxies as well as direct detection signals and upper limits, showing considerable impact in constraining the CMSSM parameters. While claimed DM signals from the CoGeNT experiment could not be fit within a CMSSM model, Xenon100 upper limits contributions have been shown.   
With upcoming fits using AstroFit and Fittino, the parameter space of DM models can be constrained even further, using latest results from the Xenon100 (as well as predictions from Xenongoal and Xenon1T) and CRESST direct detection instruments, incoming results from the LHC and indirect detection information. As numerous results in the field of experimental DM physics are expected in the near and mid-term future, it is extremely important to strengthen integrative approaches, which AstroFit facilitates.
Within AstroFit, it is planned to provide subroutines for all major observables in DM research, such as photon fluxes in different energy regimes and from different sources, antiproton, positron and neutrino fluxes in addition to the relic density of Cold Dark Matter and data on the scattering cross-section from direct detection, which can be used together with measurements from collider searches.
The inclusion of available information in particle physics, astroparticle physics and cosmology in a combined DM search is an important tool to help interprete state-of-the-art physics in these disciplines. As such, it is planned to release a public version of AstroFit for a larger community of researchers to use.

\section*{Acknowledgements}
 NN acknowledges financial support through the DFG funded collaborative research center SFB 676. Also, NN acknowlegdes the distinguished collaboration within the Fittino community with particular regards to P. Bechtle, X. Prudent and B. Sarrazin.
TB is supported by the German Research Foundation (DFG) through the Emmy Noether grant BR 3954/1-1.

\bibliographystyle{ws-procs9x6}
\bibliography{ws-pro-sample}

\end{document}